\documentclass[10pt]{iopart}
\usepackage{epsfig}
\usepackage{cite}
\usepackage{times}

\begin{document}

\noindent\fbox{\parbox{\textwidth}{
\small
Contribution to the International Workshop on Interdisciplinary 
Applications of 
Ideas from Nonextensive Statistical Mechanics and Thermodynamics, 
April 8-12, 2002, Santa Fe Institute, Santa Fe, New Mexico, USA.
To be published in M. Gell-Mann, C. Tsallis (eds.), 
{\it Nonextensive Entropy -- Interdisciplinary Applications}, 
Oxford University Press, Oxford, 2002.
}}

\title[Temperature fluctuations]{Temperature fluctuations and mixtures of equilibrium states in the
canonical ensemble}

\author{Hugo Touchette}

\address{Department of Physics and School of Computer Science, McGill
               University, Montr\'eal, Qu\'ebec, Canada H3A 2A7. 
               (htouc@cs.mcgil.ca, htouchet@alum.mit.edu)}

\begin{abstract}
It has been suggested recently that `$q$-exponential' distributions which form the 
basis of Tsallis' non-extensive thermostatistical formalism may be 
viewed as mixtures
of exponential (Gibbs) distributions characterized by a fluctuating inverse
temperature. In this paper, we revisit this idea in connection with a detailed
microscopic calculation of the energy and temperature fluctuations present in 
a finite vessel of perfect gas thermally coupled to a heat bath. We find that the
probability density related to the inverse temperature of the gas has a form similar to
a $\chi^2$ density, and that the `mixed' Gibbs distribution inferred from this density
is non-Gibbsian. These findings are compared with those obtained by a 
number of researchers who worked on mixtures of Gibbsian distributions
in the context of velocity difference measurements in turbulent fluids as well as
secondaries distributions in nuclear scattering experiments. 
\end{abstract}

\hyphenation{pano-ra-ma}

\section{Introduction}

\label{section1}

Most if not all textbooks on thermodynamics and statistical physics define
temperature as being a quantity which, contrary to other thermodynamic
observables like energy or pressure, does not admit fluctuations. Because of
that, it is somewhat surprising to see papers with the expression
`temperature fluctuations' in their titles appearing from time to time in
serious scientific journals on subjects as various as particle physics and
fluid dynamics (see, e.g., \cite
{chui1992,ching1993,ashkenazi1999,stodolsky1995}). Indeed, how can the
temperature of a system, however small, fluctuate if one defines it `as
equal to the temperature of a very large heat reservoir with which the
system is in equilibrium and in thermal contact' \cite{kittel1988}? Also, in
the case of the reservoir, how can temperature be a fluctuating parameter if
its definition requires one to assume the thermodynamic limit, i.e., to
assume that the system acting as a reservoir is composed of an infinite
number of particles or degrees of freedom? Presumably, the thermodynamic
limit should rule out any fluctuations of thermodynamic quantities like the
mean energy or the pressure, so that if temperature is related to these
quantities, how can it fluctuate?

The solution to this conundrum is quite simple. First, the standard
definition of temperature found in textbooks is too restrictive: there is
not \textit{one} but \textit{many} definitions of temperature and of
quantities analogous to temperature, as well as many physical
(non-equilibrium) situations in the context of which these different
definitions admit fluctuations \cite{mandelbrot1989}. Second, the standard
definition of temperature involving the thermodynamic limit is only an
idealization, a ``purist'' definition. Real physical bodies are always
composed of a finite number of particles or degrees of freedom, which means
that the concept of temperature must be applicable outside the idealized
realm in which it is defined if experimentalists are indeed able to measure
the temperature of real bodies in real laboratories. Physically, this means
also that there must be a threshold number of particles or degrees of
freedom (more or less precisely defined) above which a system can be
measured or ``felt'' to have a temperature \cite{feshbach1987}. Well above
this number, temperature is assured to be defined and likely to be constant,
while close to this number it may be well-defined, but may change in time or
vary in space; that is, it may fluctuate!

All the studies concerned with temperature fluctuations exploit one of the
above `indents' to the standard definition of temperature. That is, they
either consider alternatives to the standard definition of temperature which
do admit fluctuations or apply the `thermodynamic-limit' definition of
temperature in situations where the thermodynamic limit is assumed to be
``effectively'' reached without being reached ``formally'', so to speak. Our
aim in this paper is to review a number of the these alternative definitions
and situations, and to dispel, in doing so, some of the misunderstanding and
misconceptions surrounding the notion of temperature fluctuations. We will
be particularly interested in giving a detailed calculation of temperature
fluctuations present in a system which is commonly thought to be at constant
temperature, namely a system composed of a finite number of independent
particles (basically a finite volume of perfect gas) thermally coupled to an
infinite-size heat reservoir at constant temperature. In the following, we
will show that if instead of defining the temperature of the particle system
simply as being equal to the temperature of the reservoir we apply the
statistical definition of temperature to the finite-size system of particles
(provided that the number of particles is sufficiently large), then we must
come to the conclusion that the system's temperature is fluctuating, just as
its internal energy is fluctuating because of the thermal coupling with the
heat reservoir. The temperature of the particle system, in this case, is
precisely related to its internal energy, and can be seen as a
`microcanonical temperature' associated with the microscopic configurations
of a system whose internal energy is held fixed during a period of time
shorter than the energy fluctuations time scale.

Our motivation for studying the temperature fluctuations of a system of
particles in a canonical ensemble setting, and for presenting moreover this
study in a book about non-extensive statistical mechanics is threefold.
First, a system composed of a bunch of independent particles coupled to a
heat bath is one of the few thermodynamic systems for which the probability
density describing the temperature fluctuations can be calculated directly
using `first-principle' or `microscopic' arguments. Second, for this
specific system, the probability density of the temperature happens to be
very similar to a class of $\chi ^2$ densities of temperature fluctuations
recently introduced by Wilk and W\l odarczyk \cite
{wilk2000,wilk2002,wilk22002}, as well as by Beck \cite
{beck2001,beck22002,beck42002}, in the context of non-extensive statistical
mechanics \cite{tsallis1988,abe2001}. Finally, the models of non-extensive
behavior proposed by these authors are all based on the idea of `mixed
equilibrium states,' i.e., near-equilibrium states of systems characterized
by fluctuating temperatures. In the context of the present study, this idea,
as we will see, arises very naturally.

\section{Phenomenology of temperature fluctuations}

\label{section2}

One can imagine many different systems exhibiting temperature fluctuations.
The common characteristic of all of these systems is that they are
non-equilibrium systems. Below, we list and briefly comment four systems or,
more precisely, four generic situations for which temperature can be defined
and be thought to fluctuate. The list is far from being exhaustive: the
first three situations are presented to give an idea of the physical
phenomena involving temperature fluctuations which have been discussed from
the point of view of non-extensive statistical mechanics recently. The
fourth and last case of the list, the particles and heat bath system, is the
focus of this paper (see Section \ref{section3}).

\begin{itemize}
\item  \textbf{Temperature fluctuations in a gas.\ }A system with
fluctuations of temperature `spread' over space can be constructed simply in
the following way. Take a vessel of gas, and divide it in some number of
compartments thermally insulated from one another. Bring the content of each
compartment at different temperatures, and then remove the insulating
partition. From the moment where the partition is removed, a process of
temperature relaxation will take place, whereby the particles forming the
gas will collide and exchange energy until a state of uniform temperature is
achieved.

The details of the relaxation process are quite complicated at the
microscopic level, and depend on the nature and properties of the gas
considered. But, at the macroscopic level, the net result of this experiment
is simply described: between the time where the partition is removed and the
time where the gas' temperature is completely uniform, the temperature field
of the gas will vary in space as well as in time. Thus, as a whole, the
vessel of gas can be said to be in a state of fluctuating temperature. This
is admittedly an expletive way to say that the temperature is not
homogeneous in space, but the expression is nonetheless correct and widely
used (e.g., when referring to the spatial temperature fluctuations of the
cosmic background radiation).

Experimentally, there are various ways by which one can reconstruct the
temperature field of the gas, apart from plunging a thermometer into it at
different places. A simple method (conceptually, not experimentally)
consists in measuring the momenta (along a fixed direction) of many
particles of the gas at one point in space or, equivalently, sample the
momentum of a single particle over some period of time, and construct from
the measurements a histogram of the number of particles $L(x)$ having a
momentum\ value between $x$ and $x+\Delta x$. ($\Delta x$ is the
coarse-graining scale at which two particles are considered to have
different momentum values.) If the sample of measurements is large enough,
then it is expected that the form of $L(x)$ should approximately be
Gaussian, as predicted by Maxwell and Boltzmann, with a variance
proportional to the temperature $T$. Hence, fitting $L(x)$ with a Gibbs
distribution proportional to $\exp (-\beta x^2)$ or calculating its variance
or its half-width all constitute operational procedures for probing the
temperature $T=(k_B\beta )^{-1}$ of the gas. It should be noted that the
accuracy of any of these methods for obtaining $\beta $ depends on (i) the
number of measurements used to construct $L(x)$, which should be large, but
not necessarily infinite!; and (ii) the assumption that the gas is
non-interacting (perfect) or weakly interacting. These two points are
necessary to assume that the momenta of the particles are Gaussian
distributed \cite{reif1965}.

\item  \textbf{Velocity temperature in turbulent fluids.} It is common in
turbulent flow experiments to define an analog of temperature by looking at
the distribution $L(x)$ of particle velocity differences in a restricted
region of a fluid using anemometry or interferometry equipments \cite
{johnson1998}. Just as in the case of the gas, temperature is defined for a
fluid by fitting $L(x)$ with a Gibbs distribution of the form $e^{-\beta
u(x)}$, where $u(x)$ is the one-particle energy function taken to be a
quadratic or a nearly quadratic function of the velocity variable. This
defines a local inverse temperature $\beta $ which, it is important to note,
does not represent the physical inverse temperature of the fluid. Rather, it
is a correlate of the local rate of energy dissipation that takes place at
the microscopic level over a time scale known as the Kolmogorov time \cite
{ashkenazi1999,beck2001}.

The Gibbsian character of $L(x)$ and the fluctuations of the velocity
temperature in space, related to the spatial fluctuations of the local
energy dissipation rate, have been observed in many experiments of weakly
turbulent fluids (see, e.g., \cite{beck22001,arimitsu2002} and references
cited therein). However, for fluids at high Reynolds number, i.e., highly
turbulent fluids, a totally different behavior of $L(x)$ is observed.
Indeed, recent experiments have demonstrated that $L(x)$ in strong
turbulence regimes is not Gibbsian; instead, it takes the form of a
power-law which appears to be well-fitted by a so-called $q$-exponential
function 
\begin{equation}
e_q^{-\beta _qu(x)}=[1-(1-q)\beta _qu(x)]^{1/(1-q)},
\end{equation}
where $\beta _q^{-1}$ is a fitting parameter analogous to temperature \cite
{beck2001,beck22001,arimitsu2002}. To account for this non-Gibbsian
behavior, Beck has suggested to interpret $q$-exponential distributions as
`mixed' distributions arising from an ensemble of exponential distributions $
e^{-\beta u(x)}$ parameterized by a fluctuating inverse temperature $\beta $ 
\cite{beck2001,beck22002,beck42002}. That is to say, if one assumes that
what is probed in those experiments is not \textit{one} velocity
distribution $L(x)$ characterized by a fixed temperature, but a \textit{
continuum} of distributions $L(x)$ having different temperatures, then what
should be observed physically is an \textit{average} Gibbs distribution, the
average being performed over the temperature fluctuations. In this context,
the essential point made by Beck (see \cite{beck2001} for the details) is
that, if the probability density $f(\beta )$ ruling the temperature
fluctuations has the following form: 
\begin{equation}
f(\beta )=\frac 1{\Gamma \left( \frac 1{q-1}\right) }\left[ \frac
1{(q-1)\beta _0}\right] ^{\frac 1{q-1}}\beta ^{\frac 1{q-1}-1}\exp \left[
-\frac \beta {(q-1)\beta _0}\right] ,  \label{fb2}
\end{equation}
where $\beta \geq 0$ and $q>1$, then the mixed distribution obtained by
averaging the Gibbs kernel $e^{-\beta u(x)}$ with $f(\beta )$ is $q$
-exponential. Indeed, one can readily verify that 
\begin{equation}
e_q^{-\beta _0u(x)}=\int_0^\infty e^{-\beta u(x)}f(\beta )d\beta 
\label{hilhorst}
\end{equation}
using the above variant of the $\chi ^2$ or gamma density \cite{freund1992}
for $f(\beta )$. This integral representation of the $q$-exponential
function is sometimes referred to as Hilhorst's formula \cite
{tsallis1994,prato1995}.

\item  \textbf{Nuclear collision temperature.} The basic idea involved in
the definition of temperature in nuclear scattering experiments is to
consider the set of particles produced during a collision (called the
products) as forming a gas of particles which, at a first level of
approximation, can be treated as being non-interacting (perfect gas
approximation). From this point of view, a concept of `collision
temperature' is defined essentially in the same way that temperature was
defined for turbulent fluids except that the precise physical property to
look at in scattering experiments is not the shape of the momenta
distribution itself, but the so-called exponential dependence of the
distribution of secondaries with respect to transverse momentum \cite
{stodolsky1995,beck42002}.

Since the number of particles probed during one scattering experiment is
never very large ($\sim 10-1000$), one must sometimes collect the momenta of
particles over many scattering experiments before the exponential shape of
the secondaries distribution reveals itself. However, this is not always the
case:\ in heavy-ion experiments at very high energy, for example, it is
often observed that a single event, i.e., only one scattering experiment is
sufficient for a thermostatistical analysis to be effective \cite
{stodolsky1995}. Also, what is often seen is that scattering events of same
nature repeated over time yield different collision temperatures, making
obvious that temperature is a fluctuating parameter.

Observations of `non-extensive' behavior in relation to this thermodynamic
picture of scattering experiments have been reported so far on two different
fronts. The first is related to the distribution of secondaries, and, more
precisely, to observed deviations of this distribution from its expected
exponential form. Due to the limited space available here, we will not
discuss this case as it is quite involved. Let us only mention that Wilk and
W\l odarczyk have advanced in \cite{wilk2002} a `mixed exponential
distribution' model of these deviations analogous to the one suggested by
Beck.

The second case of `non-extensive' behavior concerns the absorption of
cosmic ray particles in lead chambers \cite{wilk2000,wilk2002,wilk22002}.
This case was also studied by Wilk and W\l odarczyk who suggested for its
explanation yet another variant of the $\chi ^2$ temperature fluctuations
model (actually before Beck applied similar ideas to the study of turbulent
fluids). The physics explained by their model is the following. The number $N
$ of hadronic particles absorbed in lead chambers is usually measured to be
distributed as a function of the depth $l$ according to 
\begin{equation}
\frac{dN}{dl}\propto e^{-l/\lambda },
\end{equation}
where $\lambda $ is the mean free path parameter or mean penetration depth
(an analog of temperature). This exponential distribution is, at least, what
is observed at small penetration depths ($\sim 60$ cm of lead); beyond that,
what is observed is that $dN/dl$ changes to a power-law which can be fitted
by a $q$-exponential with $q\simeq 1.3$. To account for this crossover, Wilk 
\textit{et al.}~simply conjectured that the $\lambda $ parameter
characterizing the long flying components (i.e., the deep penetration
events) is subject to fluctuations, and, thus, that the $q$-exponential
penetration profiles observed experimentally for these components are
mixtures of exponential distributions. By assuming that the probability
density of $\lambda $ is a $\chi ^2$ density, they were effectively able to
reproduce the non-exponential distributions measured in laboratories \cite
{wilk2000,wilk2002,korus2001}.

\item  \textbf{System coupled to a heat bath.}\ Our last example in the
panorama of thermodynamic systems characterized by temperature fluctuations
is the prototypical system defining the canonical ensemble: that is, a small
system $S$ in thermal contact with a larger system $R$ acting as a heat
reservoir. Following the standard textbook definition of the canonical
ensemble, one should say that the temperature of system $S$ at equilibrium
is constant, and is equal to the temperature of system $R$; after all, this
is how thermal equilibrium is defined. However, such a statement does not do
justice to one important property of $S$ which is that the energy density of 
$S$ fluctuates (because of its finiteness) while the energy density of $R$
does not (by definition of a heat bath).

To make this statement more precise, suppose that $S$ consists of $n$
independent particles whose energy density or mean total energy is given by 
\begin{equation}
U_n=\frac 1n\sum_{i=1}^nu_i.
\end{equation}
Since the particles are coupled to $R$, the $u_i$'s above are random
variables, which means that $U_n$ is also a random variable. Moreover,
observe that $U_n$, for any finite $n$, has a non-negligible probability to
assume many different values because, in this case, the probability density $
g_n(u)$ of $U_n$ is not a Dirac-delta function. The Dirac density, formally,
is only a limiting density which ``attracts'' $g_n$ as $n\rightarrow \infty $
. (This basically follows from the law of large numbers.) Thus, if we can
associate an inverse temperature $\beta (u)$ to all energy states such that $
U_n=u$, e.g., by applying the equipartition theorem or by fitting a
distribution of energy levels with a Gibbs distribution as described
earlier, then we must conclude that there are different values of $\beta $
effectively realized `in' or `by' the particle system, so to speak. That is
to say, the probability density $f_n(\beta )$ for $\beta $, obtained from $
g_n(u)$ by a change of variables $u\rightarrow \beta (u)$, cannot be a
Dirac-delta function if $g_n(u)$ is not itself a delta function. It is to be
expected that $f_n(\beta )\rightarrow \delta (\beta -\beta _0)$, where $
\beta _0$ is the inverse temperature of heat bath, only in the thermodynamic
limit where $n\rightarrow \infty $. These points are discussed in more
mathematical details in the next sections.
\end{itemize}

\section{Energy and temperature fluctuations in the canonical ensemble}

\label{section3}

Our analysis of energy and temperature fluctuations of a system coupled to a
heat bath will be presented in the context of the following model. Let a
vessel of gas containing $n$ independent (classical) particles be thermally
coupled to a heat reservoir characterized by a fixed inverse temperature $
\beta _0$. The state of each particle is represented by a random variable $
X_i$, $i=1,2,\ldots ,n$, to which is associated a (one-particle) energy $
u(X_i)$. The set of outcomes of each of the $X_i$'s (the one-particle state
space)\ is denoted by $\mathcal{X}$. With these notations, the energy
density or mean energy of the gas is written as 
\begin{equation}
U_n(x^n)=\frac 1n\sum_{i=1}^nu(x_i)=\sum_{x\in \mathcal{X}}L_n(x)u(x),
\label{energ1}
\end{equation}
where $x^n=x_1,x_2,\ldots ,x_n$ is the joint state of the system, i.e., the
state of the system as a whole. Note that in the above expression we have
defined $L_n(x)$ as the relative number of particles which are in state $x$,
i.e., as 
\begin{equation}
L_n(x)=\frac{\#( \mathrm{particles} :X_i=x)}n.
\end{equation}
It should be noted that the vector $L_n$ is nothing but the histogram of
one-particle states referred to as previously when we discussed temperature
fluctuations. Indeed, in the case where $x$ represents a momentum variable,
the quantity $nL(x)$ precisely counts the number of particles having a
momentum value equal to $x$. (We assume throughout that the $X_i$'s are
discrete random variables; the continuous case can be treated with minor
modifications.)

Now, owing to the fact that the gas is treated in the canonical ensemble, in
the sense that it is coupled to a heat bath, we have 
\begin{equation}
P_n(x^n)=P_n(X_1=x_1,X_2=x_2,\ldots ,X_n=x_n)=\frac{e^{-\beta _0nU_n(x^n)}}{
Z_n(\beta _0)}  \label{pcan1}
\end{equation}
as the joint probability distribution over the states $x^n$, where 
\begin{equation}
Z_n(\beta _0)=\sum_{x^n\in \mathcal{X}^n}e^{-\beta _0nU_n(x^n)}
\end{equation}
is the $n$-particle partition function. Of course, since all the particles
are assumed to be independent (perfect gas assumption), as well as all
individually coupled to the same heat bath, we can also write 
\begin{equation}
P_n(x^n)=p(x_1)p(x_2)\cdots p(x_n)=\frac{e^{-\beta _0u(x_1)}}{Z(\beta _0)} 
\frac{e^{-\beta _0u(x_2)}}{Z(\beta _0)}\cdots \frac{e^{-\beta _0u(x_n)}}{
Z(\beta _0)}  \label{pcan2}
\end{equation}
with $Z(\beta _0)=Z_1(\beta _0)$ (one-particle partition function). These
equations make obvious the fact that what we are dealing with is a system of
independent and identically distributed (IID) random variables.

The first quantity that we are interested to calculate at this point is the
probability distribution or probability density $g_n(u)$ associated with the
outcomes $U_n=u$. \textit{A priori}, finding an exact expression for $g_n(u)$
is not an easy task, even though $U_n$ is the simplest sum of random
variables that one can imagine, i.e., one involving IID random variables.
Fortunately, there exists a general method by which one can obtain a very
accurate approximation of $g_n(u)$ for $n\gg 1$ without too much efforts.
This method is based on the theory of large deviations \cite{dembo1998}, and
proceeds by observing that probability densities of normalized sums of IID
random variables, such as the one defining $U_n$, satisfy two basic
properties: (i) they decay exponentially with the number $n$ of random
variables involved; and (ii) the rate of decay is a function of the value
(outcome) of the sum alone. In the present context, this means specifically
that $g_n(u)$ has the form 
\begin{equation}
g_n(u)\asymp e^{-nD(u)}.
\end{equation}
The sign `$\asymp $' above is there to emphasize that the large deviation
approximation of the density $g_n(u)$ is `exponentially tight' with $n$,
i.e., that it is exact up to $O(n^{-1}\ln n)$ marginal corrections to the
rate of decay $D(u)$. This rate of decay or \textit{rate function} is itself
calculated as the Legendre transform of the quantity 
\begin{equation}
\lambda (k)=\ln E[e^{ku(X)}]=\ln \sum_{x\in \mathcal{X}}p(x)e^{ku(x)}
\end{equation}
which is the \textit{cumulant generating function} of the probability
distribution $p(x)$ associated with the IID\ random variables. The result of
this transform is 
\begin{equation}
D(u)=uk(u)-\lambda (k(u)),  \label{rate1}
\end{equation}
$k(u)$ being the solution of 
\begin{equation}
\left. \frac{d\lambda (k)}{dk}\right| _{k(u)}=u.  \label{leg1}
\end{equation}
A proof of this result can be found in \cite
{dembo1998,oono1989,ellis1995,ellis1999} (see also the notes contained in 
\cite{dembo1998} for a historical account of the developments of the theory
of large deviations together with a list of the founding papers of this
theory).

Physicists who are not familiar with the formalism of large deviations will
probably look at the above formulae for calculating $g_n(u)$ as being quite
formal if not fancy (in a pejorative way). For them, we offer the following
alternative derivation of $g_n(u)$. Consider the density $\Omega _n(u)$ of
states $x^n$ having the same energy $U_n(x^n)=u$. Following the
thermostatistics of Gibbs and Boltzmann, this density of states must be an
exponential function of $n$ taking the form 
\begin{equation}
\Omega _n(u)\asymp e^{nH(u)},  \label{dens1}
\end{equation}
where $H(u)$ is the entropy of the system at energy density $u$. As is
well-known, the function $H(u)$ is also obtained by a Legendre transform,
this time involving the logarithm of the one-particle partition function or
free energy. Now, using the above approximation for $\Omega _n(u)$, and the
fact that all states $x^n$ such that $U_n(x^n)=u$ have the same probability 
\begin{equation}
P_n(x^n:U_n(x^n)=u)=\frac{e^{-\beta _0nu}}{Z(\beta _0)},
\end{equation}
we can write 
\begin{equation}
g_n(u)=\Omega _n(u)P_n(x^n:U_n(x^n)=u)\asymp e^{-n[u\beta _0+\ln Z(\beta
_0)-H(u)]}.
\end{equation}
Thus, we arrive at 
\begin{equation}
D(u)=u\beta _0+\ln Z(\beta _0)-H(u).  \label{rate2}
\end{equation}
One can verify that the above expression for the rate function is totally
equivalent to the one found in the context of large deviation theory. Both
expressions are, in fact, related by the transformation $\beta (u)=\beta
_0-k(u)$, where 
\begin{equation}
\beta (u)=\frac{dH(u)}{du}  \label{b1}
\end{equation}
is the usual thermostatistical definition of the inverse temperature. The
proof of this equivalence result follows, essentially, by noting that 
\begin{equation}
\lambda (k)=\ln \sum_{x\in \mathcal{X}}\frac{e^{-\beta _0u(x)}}{Z(\beta _0)}
e^{ku(x)}=\ln Z(\beta _0-k)-\ln Z(\beta _0),
\end{equation}
and by using the familiar expression $H(u)=u\beta (u)+\ln Z(\beta (u))$ for
the entropy. (The complete verification of the result is left as an exercise
to the reader.)

Let us now turn to the matter of defining an inverse temperature $\beta $
for our system of IID particles, and to the complement matter of inferring
the probability density $f_n(\beta )$. Following our discussion of
temperature fluctuations, it should be expected that there are many ways by
which one can assign a temperature to the microcanonical set of states
defined by 
\begin{equation}
M_n(u)=\{x^n:U_n(x^n)=u\}.
\end{equation}
Also, it is to be expected that one definition of temperature may not
necessarily coincide with another in the case of finite-size ($n<\infty $)
systems. We illustrate this possibility by comparing below four different
definitions or `flavors' of temperature.

\begin{itemize}
\item  \textbf{Derivative of entropy or free energy.} An obvious way to
associate a temperature to the states in $M(u)$ is to take the energy
derivative of the microcanonical entropy $H(u)$ as in Eq.(\ref{b1}).
Equivalently, one can solve the equation 
\begin{equation}
-\frac{d\ln Z(\beta )}{d\beta }=u  \label{part1}
\end{equation}
for $\beta $, or compute the function $k(u)$ from Eq.(\ref{leg1}) and use
the relation $\beta (u)=\beta _0-k(u)$. The inverse temperature obtained by
any of these methods will be denoted by $\beta _{th}(u)$ to emphasize that
it is based on \textit{intensive} thermodynamic potentials which do not
depend on $n$.

\item  \textbf{Derivative of the density of state.} A slightly different
definition of inverse temperature is obtained by taking the `logarithmic
derivative' of $\Omega _n(u)$ with respect to the total energy $nu$ 
\begin{equation}
\beta _\Omega (u)=\frac 1{\Omega _n(u)}\frac{d\Omega _n(u)}{d(nu)}=\frac{
d\ln \Omega _n(u)}{d(nu)}
\end{equation}
\textit{in lieu} of the derivative of the entropy exponent as in Eq.(\ref{b1}
). This defines another inverse temperature $\beta _\Omega (u)$ which
differs from $\beta _{th}(u)$ by a term of order $O(n^{-1}\ln n)$ which
vanishes as $n\rightarrow \infty $.

\item  \textbf{Gibbsian distribution of states.} An inverse temperature $
\beta _L(u)$ can be defined from a phenomenological point of view by fitting
a given distribution of states $L_n(x)$ of mean energy $U_n=u$ with a Gibbs
distribution of the form 
\begin{equation}
L^u(x)=\frac{e^{-\beta _L(u)u(x)}}{Z(\beta _L(u))}.
\end{equation}
We have described this definition of temperature earlier (see Section \ref
{section2}), and have noted that it is accurate when $n$ is large. To be
more precise, it is accurate in a probabilistic sense because, in theory,
there is always a possibility that non-Gibbsian distributions $L_n(x)$ of
mean energy $U_n=u$ can be observed. However, the probability associated
with such a possibility is very small and vanishes rapidly as $n\rightarrow
\infty $. To see why, let us consider all the states $x^n$ and their
corresponding distributions $L_n$ present in the energy `box' $M(u)$. What
we want to show is that the probability $P_n(L^u)$ that $L^u$ is observed in 
$M(u)$ is overwhelmingly large compared to the probability $P_n(L)$ to
observe any other distribution $L\neq L^u$. To show this, we use another
result of the theory of large deviations \cite{oono1989,ellis1995,ellis1999}
which states that 
\begin{equation}
\frac{P_n(L)}{P_n(L^u)}\asymp e^{-n[H(L^u)-H(L)]}=e^{-n\Delta H},
\label{sanov1}
\end{equation}
where 
\begin{equation}
H(L)=-\sum_{x\in \mathcal{X}}L(x)\ln L(x)  \label{sanov2}
\end{equation}
is the Boltzmann-Gibbs-Shannon entropy, and $\Delta H=H(L^u)-H(L)$. Using
the fact that $L^u$ is a maximum entropy distribution under the constraint $
U_n=u$, it is easy to see that $\Delta H\geq 0$ with equality if and only if 
$L=L^u$, so that $P_n(L)/P_n(L^u)\rightarrow 0$ as $n\rightarrow 0$.
Moreover, the discrepancy between the two probabilities is exponentially
large in $n$. Thus, for $n$ large it can be said that any distribution $L_n$
picked at random in $M(u)$ will be such that $L_n\simeq L^u$. As this holds
for any $M(u)$, this implies that any measured distribution related to some
randomly chosen state $x^n$ with $n\gg 1$ ought to be a Gibbs distribution
or be very close to a Gibbs distribution with a probability nearly equal to
1.
\end{itemize}

The preceding paragraphs show that there is some arbitrariness in defining
the concept of temperature for systems composed of a finite number of
particles or degrees of freedom. In theory, there is some indeed; however,
if $n$ is large, then defining the temperature in any of the ways described
above should have little effect on the actual value of the temperature
inferred. Thus, for all practical purposes, we can assume that $\beta
_{th}(u)\simeq \beta _\Omega (u)\simeq \beta _L(u)$ for $n\gg 1$. In view of
what was said in Section \ref{section2}, it should be noted that the
particular approximation $\beta _{th}(u)\simeq \beta _L(u)$ is of deep
consequences: if we look at the distributions $L_n(x)$ associated to the
states $x^n\in \mathcal{X}^n$, then we are likely to realize that the
majority of these distributions, i.e., those which have the most probability
to be observed, form a set of Gibbs distributions $L^u$ parameterized by a
fluctuating inverse temperature $\beta (u)$ (from now on we do not
distinguish between the different flavors of inverse temperature). This
means that for $1\ll n<\infty $ all the statistical and thermodynamic
properties of our system can be described, in an effective manner, using an
ensemble of Gibbs distributions with a fluctuating temperature. The
probability density $f_n(\beta )$ ruling the inverse temperature
fluctuations must, in this case, be given by 
\begin{equation}
f_n(\beta )=g_n(u(\beta ))\left| \frac{du(\beta )}{d\beta }\right| ,
\label{temp2}
\end{equation}
where $u(\beta )$ is the inverse function of $\beta (u)$. Using this
density, one then defines a `mixed' or `average' Gibbs distribution of
one-particle states as follows: 
\begin{equation}
\tilde{L}(x)=\int L^{u(\beta )}f_n(\beta )d\beta =\int \frac{e^{-\beta u(x)} 
}{Z(\beta )}f_n(\beta )d\beta .  \label{mix1}
\end{equation}
This integral is a definite integral which must be evaluated over the range
of definition of $\beta $. Equivalently, the average can be taken over the
energy coordinate: 
\begin{equation}
\tilde{L}(x)=\int L^ug_n(u)du=\int \frac{e^{-\beta (u)u(x)}}{Z(\beta (u))}
g_n(u)du.
\end{equation}
In the above equation, be sure to distinguish the energy function $u(x)$
from the value $u$ of the mean energy $U_n$. Also note the slight difference
between these mixed distributions and those proposed by Wilk \textit{et al.}
~and Beck: in our version of mixed distributions, we take the average over
the Gibbs factor $e^{-\beta u(x)}$ \textit{normalized} by the partition
function which is itself a function of $\beta $ (compare Eqs.(\ref{hilhorst}
) and (\ref{mix1})).

\section{The case of the perfect gas}

\label{section4}

As an application of the large deviation formalism, we carry out in this
section the complete calculation of $g_n(u)$ and $f_n(\beta )$ for $
u(x)=x^2/2$. By using this form of energy, we assume that the particles
composing the gas have a unit mass, and that their momentum $x_i$, $
i=1,2,\ldots ,n$, is confined to one dimension ($\mathcal{X}$ is the real
line extending from $-\infty $ to $+\infty $). We also abstract out the
position of the particles from the analysis, since the mean energy $U_n$ of
the $n$ particles does not depend on the position degree of freedom.

To find $g_n(u)$, we first calculate the rate function $D(u)$ using the
Legendre transform method. The cumulant generating function associated with
the quadratic energy function is calculated to be 
\begin{equation}
\lambda (k)=\ln \int_{-\infty }^\infty \frac{e^{-\beta _0x^2/2}}{Z(\beta _0)}
e^{kx^2/2}dx=\frac 12\ln \frac{\beta _0}{\beta _0-k}.
\end{equation}
From this equation, we find the `translated' inverse temperature $k(u)$ by
solving 
\begin{equation}
\left. \frac{d\lambda (k)}{dk}\right| _{k(u)}=\frac 12\frac 1{\beta
_0-k(u)}=u
\end{equation}
The solution is $k(u)=\beta _0-(2u)^{-1}$, so that 
\begin{equation}
D(u)=uk(u)-\lambda (k(u))=u\beta _0-\frac 12-\frac 12\ln 2\beta _0u.
\end{equation}
Thus, 
\begin{equation}
g_n(u)\asymp u^{n/2}e^{-nu\beta _0}.  \label{gn1}
\end{equation}
This form of density is a variant of the $\chi ^2$ or gamma density
mentioned previously with $n$ as the number of degrees of freedom \cite
{freund1992}. Note that this density for the mean energy can be derived
directly by noting that $U_n$, for $u(x)=x^2/2$, is a normalized sum of
squares of $n$ Gaussian random variables. In statistics, this is usually how
the $\chi ^2$ density is introduced \cite{freund1992}.

\begin{figure}[t]
\begin{center}
\epsfig{file=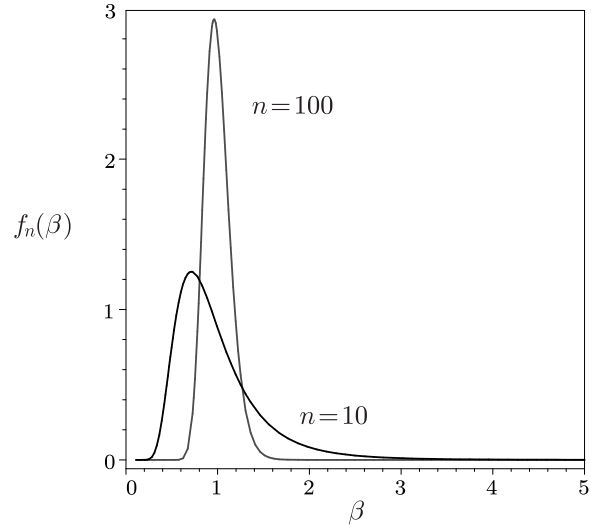,clip=}
\end{center}
\caption{Probability densities $f_n^{ld}(\beta)$ characterizing the $\beta$
fluctuations of $n=10$ and $n=100$ free particles thermally coupled to a
heat bath with $\beta_0=1$. Each density is defined for $\beta>0$, and shows
a maximum at $\beta_0n/(4+n)$.}
\label{figure1}
\end{figure}

At this point, the density $f_n(\beta )$ describing the fluctuations of $
\beta$ is readily deduced from the expression of $g_n(u)$ found above by
calculating the physical inverse temperature $\beta (u)$. To this end, we
can use the fact that $\beta (u)=\beta _0-k(u)$ or use the equipartition
theorem to find in both cases that $\beta (u)=(2u)^{-1}$. Hence, following
Eq.(\ref{temp2}), $f_n(\beta )$ must have the form 
\begin{equation}
f_n(\beta )\asymp \frac 1{\beta ^{n/2}}e^{-\frac{n\beta _0}{2\beta }}\frac
1{\beta ^2}.
\end{equation}
Normalizing this expression for $\beta \in [0,\infty )$ yields 
\begin{equation}
f_n^{ld}(\beta )=\frac{\beta _0}{\Gamma (\frac n2)}\left( \frac{n\beta _0}
2\right) ^{n/2}\beta ^{-n/2-2}e^{-\frac{n\beta _0}{2\beta }}
\end{equation}
as the large deviation ($ld$) approximation of $f_n(\beta )$. A plot of this
density for two values of $n$ ($10$ and $100$) is shown in Fig.~\ref{figure1}
with $\beta _0=1$. The plot corresponding to $n=10$ has no real physical
significance, since the large deviation approximation is not expected to be
effective in this case. However, it is presented to illustrate the skewness
(to the right) of $f_n(\beta )$ which disappears as $n\rightarrow \infty$.
The maximum value of $f_n(\beta )$ is given by $\beta _{\max }=\beta
_0n/(4+n)$. As expected, $f_n(\beta )$ converges (in a uniform sense) to the
thermodynamic-limit density $f_\infty (\beta )=\delta (\beta -\beta _0)$
when $n\rightarrow \infty $; this is partially seen by looking at Fig.~\ref
{figure1}. By virtue of the law of large numbers, $g_n(u)$ must also
converge in the same limit to a $\delta $ density taking this time the form $
g_\infty (u)=\delta (u-u(\beta _0))$ where 
\begin{equation}
u(\beta _0)=E[u(X)]=\int_{-\infty }^\infty \frac{e^{-\beta _0x^2/2}}{Z(\beta
_0)}\frac{x^2}2dx=\frac 1{2\beta _0}.
\end{equation}

\begin{figure}[t]
\begin{center}
\epsfig{file=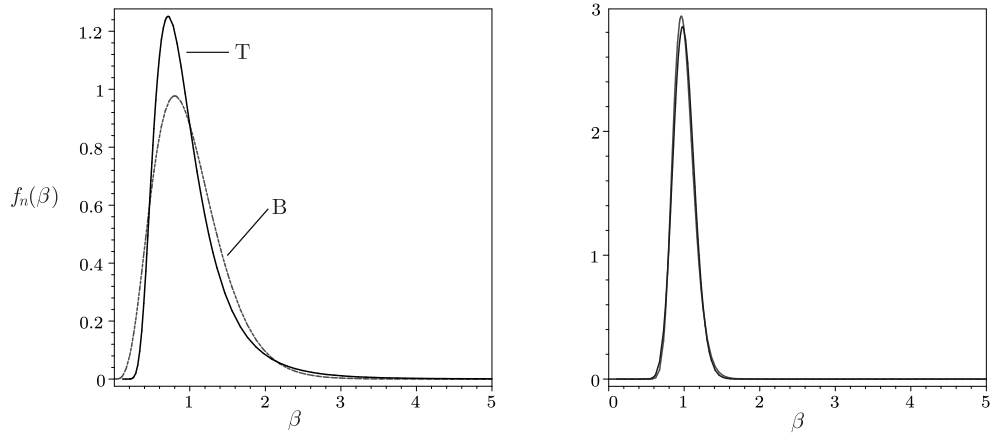,width=\textwidth,clip=}
\end{center}
\caption{Comparison of the $\chi^2$ $\beta$-density proposed by Beck (B) 
\protect\cite{beck2001} and the one proposed in this work (T) for $n=10$
(left) and $n=100$ (right) particles (see text). For $n=100$, the two
densities are quasi-indistinguishable. The maximum value of the $\beta$
-density, in the case of Beck, is located at $\beta_0(n-2)/n$.}
\label{figure2}
\end{figure}

We now come to the main point of our study which is to compare the $f_n(\beta )$
density obtained here and the $\chi ^2$ probability density of Eq.(\ref{fb2}
) which was `postulated' by Wilk \textit{et al.}~and Beck in their studies
of mixed distributions (see Section \ref{section2}). To establish this
comparison, we present in Fig.~\ref{figure2} two plots of $f_n^{ld}(\beta )$
for two different values of $n$ and a variant of the $\chi ^2$ $\beta $
-density proposed by Beck 
\begin{equation}
f_n^B(\beta )=\frac 1{\Gamma (\frac n2)}\left( \frac n{2\beta _0}\right)
^{n/2}\beta ^{n/2-1}e^{-\frac{n\beta }{2\beta _0}},
\end{equation}
which results from identifying $1/(q-1)$ in Eq.(\ref{fb2}) with $n/2$ \cite
{beck2001}. The plots are presented again for $n=10$ and $n=100$. A rapid
inspection of the expressions of $f_n^{ld}$ and $f_n^B$ reveals that these
densities are not at all the same. The density $f_n^B$ can in fact be viewed
as emerging from the sum of the squares of $n$ Gaussian random variables,
whereas, in our case, the sum of squared Gaussian random variables arises as
the mean energy, and so as $\beta ^{-1}$ modulo some constant. This explains
why performing the change of variables $\beta \rightarrow \beta ^{-1}$ in $
f_n^{ld}$ yields $f_n^B$ modulo some constant and a Jacobian term arising
from the change of variables. In spite of this important difference, the
second plot of Fig.~\ref{figure2} shows that both densities are remarkably
similar as $n$ gets large. This, at first, does not seem surprising as both
densities converge to the delta density $f_\infty (\beta )$ in the
thermodynamic limit $n\rightarrow \infty $. However, it is to be noted that
each of them gives totally different mixed distributions when they are used
in Eq.(\ref{mix1}). Indeed, it can be shown \cite{touchette2002} that the
mixed distribution associated with $f_n^{ld}$ has the asymptotic form 
\begin{equation}
\tilde{L}_n^{ld}(x)\sim e^{-\left| x\right| }
\end{equation}
for $|x|\gg 1$, instead of 
\begin{equation}
\tilde{L}^B(x)\propto e_q^{-x^2},
\end{equation}
where $q=1-2/n$. Both of these results should be compared with the (pure)
Gibbsian distribution 
\begin{equation}
L^G(x)\propto e^{-x^2}
\end{equation}
which is the limiting distribution of $\tilde{L}_n^{dl}$ and $\tilde{L}_n^B$
in the thermodynamic limit ($n\rightarrow \infty $).

The above scaling relationships clearly indicate that choosing between $
f_n^{ld}(\beta )$ or $f_n^B(\beta )$ has a dramatic consequence on the
functional form of the mixed distribution calculated even for $n\gg 1$. Does
that imply that our model of temperature fluctuations cannot serve as a
model of the non-Gibbsian distributions which have been observed in
turbulent fluid experiments as well as in nuclear scattering experiments?
The answer is not as straightforward as one would think. First, it is not at
all clear that turbulent fluids can actually be treated in the canonical
ensemble and/or that the perfect gas assumption is a valid approximation in
this case. These points call for further justifications. Second, the extreme
events $\left| x\right| \gg 1$ needed to validate either one of two
temperature densities compared in this work are often very difficult to
detect experimentally in a reliable way. Surely, additional calculations and
experimental data would be welcome in order to test the validity of our
approach, and to confront it with that of the authors mentioned in the
present study. This seems to be especially true for nuclear scattering
experiments which are usually thought to fit perfectly well into the
canonical ensemble picture \cite{stodolsky1995,cottinghman2001}.

\section{Concluding remarks}

\label{section5}

Our treatment of the energy and temperature fluctuations of a system coupled
to a heat bath has focused mainly on the perfect gas. However, it is worth
noting that the large deviation approach presented in this paper for
calculating the energy and temperature probability densities in the
canonical ensemble is very general. It can be applied independently of the
form of the energy function $u(x)$ which defines the mean energy $U_n$, and
can also be generalized without too much difficulties to cases involving
other forms of probability distribution for $P_n(x^n)$ (e.g., $q$
-exponential distributions). In this context, an obvious extension of our
work could be to consider different forms for $u(x)$, and to look at the
mixed distributions which result from the corresponding temperature
fluctuations. This line of thought has been followed recently by Beck and
Cohen \cite{beck32002} who derived a number of `superstatistical' mixed
distributions (sometimes unphysical ones) by assuming different forms of
temperature fluctuations. Another problem could be to solve the following
`inverse problem': for which $u(x)$ is $f_n(\beta )$ the same $\chi ^2$
density as the one suggested by Beck? Finally, note that a large deviation
calculation of $g_n(u)$ and $f_n(\beta )$ can also be carried out for
systems involving dependent random variables. Unfortunately, the
calculations leading to the specific forms of $g_n(u)$ and $f_n(\beta )$ in
this case are likely to be tedious. Also, the concept of mixed distribution
does not generalize easily to the case of interacting particles because the
Eqs.(\ref{sanov1}) and (\ref{sanov2}) which were used to prove that Gibbs
distributions are the only distributions likely to be observed in large
systems are valid for sequences of IID\ random variables only. It is, in
fact, a long-standing open problem of large deviation theory to generalize
these equations to sequences of dependent random variables. Solving this
problem would have direct consequences in statistical physics, for it
implies \textit{ipso facto} a generalization of the maximum entropy
principle to systems of interacting particles.

\section*{Acknowledgments}

It is a pleasure for me to thank M. Gell-Mann and C. Tsallis for the kind
invitation to participate to the Workshop on Interdisciplinary Applications
of Nonextensive Statistical Mechanics. I also want to thank S. Lloyd who
shares a great part of responsibility for my visit at the Santa Fe
Institute. This work has been supported in part by the Natural Sciences and
Engineering Research Council of Canada, and the Fonds qu\'{e}b\'{e}cois de
la recherche sur la nature et les technologies.

\end{document}